\documentclass[a4paper,11pt]{article}
\usepackage{pos}

\usepackage{bigints}

\newcommand{\mcTau}{\mathcal{T}}
\newcommand{\mcM}{\mathcal{M}}
\newcommand{\YMconfs}{\mathcal{C}_{G}\big(\mcTau\big)}
\newcommand{\YMconfsM}{\mathcal{C}_{G}\big(\mcM\big)}

\title{Coupling Yang–Mills with Causal Dynamical Triangulations}

\author[a]{Alessandro Candido}
\author*[b,c,1]{Giuseppe Clemente}
\author[d]{Massimo D'Elia}
\author[e]{Federico Rottoli}

\affiliation[a]{Dipartimento di Fisica dell’Università degli Studi di Milano and INFN 
- Sezione di Milano,\\ Via Giovanni Celoria, 16 {-} 20133 Milan, Italy}
\affiliation[b]{Institute for Mathematics, Astrophysics and Particle Physics (IMAPP)\\
    Radboud University Nijmegen, Heyendaalseweg 135, 6525 AJ Nijmegen, The Netherlands}
\affiliation[c]{Deutsches Elektronen-Synchrotron DESY, Platanenallee 6, 15738 Zeuthen}
\affiliation[d]{Dipartimento di Fisica dell'Universit\`a di Pisa and INFN
- Sezione di Pisa,\\ Largo Pontecorvo 3, I-56127 Pisa, Italy.}
\affiliation[e]{SISSA {--} International School for Advanced Studies, \\
via Bonomea 265,  34136 Trieste, Italy.}

\note{c is the current speaker affiliation}

\emailAdd{alessandro.candido@mi.infn.it}
\emailAdd{giuseppe.clemente@desy.de}
\emailAdd{massimo.delia@unipi.it}
\emailAdd{frottoli@sissa.it}

\abstract{We discuss the algorithmic problem of 
minimal coupling gauge fields of the Yang–Mills type 
to Quantum Gravity in the approach known as Causal Dynamical Triangulations (CDT) as a
step towards studying, ultimately, systems of gravity coupled with bosonic and fermionic matter. 
We first describe the algorithm for general dimensions and gauge groups 
and then focus on the results obtained from simulations of 2d CDT coupled 
to gauge fields with U{(1)} and SU{(2)} gauge groups, 
where we studied both observables related to gravity and gauge fields, 
and compared them with analogous simulations in the static flat case.}

\FullConference{%
 The 38th International Symposium on Lattice Field Theory, LATTICE2021
  26th-30th July, 2021
  Zoom/Gather@Massachusetts Institute of Technology
}

\begin{document}
\maketitle

\section{Introduction}

The asymptotic safety scenario, envisioned by Weinberg in the seventies~\cite{ass_weinberg}, 
is one of the possible solutions to the problem 
of building a self-consistent Quantum Theory of gravity. 
Grounded upon the Wilsonian approach to the renormalization group,
its key idea provides the possibility to renormalize a theory 
around a non-perturbative (non-Gaussian) UV fixed point in the space of parameters.
This scenario cannot be explored using standard perturbative techniques, 
which fail for gravity~\cite{sagnotti}, 
but must be investigated using non-perturbative techniques 
as functional renormalization group~\cite{frg_reuter} or path integral Monte Carlo methods.
Our following discussion is specialized to the latter class, in particular 
to the approach known as Causal Dynamical Triangulations 
(CDT)~\cite{cdt_pioneer,cdt_review12,cdt_review19,DynTriLor},
where the path integral is a sum over spacetime geometries approximated 
by simplicial manifolds in the so-called Regge formalism~\cite{regge}.
What distinguishes CDT from the earlier Euclidean Dynamical Triangulations approach is
the presence of an additional condition of global hyperbolicity 
by means of a space-time foliation~\cite{causconds}, 
condition that allows to consistently perform a Wick rotation from the Lorentzian 
to the Euclidean signature and that seems crucial for the appearance 
of second-order lines in the phase diagram~\cite{cdt_secondord,cdt_secondfirst,new_phase_chars,cdt_newhightrans,cdt_toroidal,cdt_phasestruct_toroidal,LBseminal,LBrunning}.

Beside pure-gauge gravity, it is worthwhile to investigate also 
the coupling of matter and gauge fields to CDT for at least two reasons:
on one hand, it will provide a dictionary for matching numerical results with 
quantum cosmological phenomenology, 
on the other hand, investigating the existence and stability of the UV non-trivial fixed point 
under different combinations of matter and gauge fields is crucial to characterize 
all possible theories in the continuum.

In this proceeding, we briefly review the results obtained in~\cite{CDTYM}, 
where we discuss the coupling of either Abelian or non-Abelian gauge theories 
to CDT and show numerical results for the gauge groups $U(1)$ and $SU(2)$.
In Section~\ref{sec:mincoup} we describe the discretization of the action adopted;
some details about the algorithmic implementation of the Monte Carlo simulation 
are discussed in Section~\ref{sec:algo}, while in Section~\ref{sec:numres} 
we present some numerical results. Finally, in Section~\ref{sec:conclusions}
we discuss the results and future developments. 

\section{Minimal coupling of Yang-Mills theories to CDT}\label{sec:mincoup}
First of all, we briefly recall 
the continuum Yang-Mills action on a flat (Euclidean) space
\begin{equation}\label{eq:un-YMact_flat}
    S_{\text{YM}} = \frac{1}{4} \bigintssss\limits_{\mathcal{M}} \! d^d x \; 
    F^{a}_{\mu \nu} F^{a \mu \nu},
\end{equation}
where $F^{a}_{\mu \nu}=\partial_\mu A^{a}_{\nu} - \partial_\nu A^{a}_{\mu} + g f^{abc} A^{b}_{\mu} A^{c}_{\nu}$,
and $A_{\mu}=A^{a}_{\mu} T^{a}$, with $T^{a}$ the generators of the Lie algebra. 
The lattice discretization usually adopted for the theory in~\eqref{eq:un-YMact_flat} 
consists of associating gauge link variables $U_\mu(n)$,
with values in the gauge group $G$, to each edge $({\bf n},{\bf n}+\hat{\mu})$ 
of a flat hypercubic lattice; these can be interpreted as elementary parallel transporters 
linking adjacent sites. 
There is some freedom also in the choice of a specific discretization for the action, 
the simplest one being the so-called \emph{plaquette} (or Wilson) action:
\begin{equation}\label{eq:un-YMact}
S_{\text{YM}} \equiv - \frac{2N}{g^2} \sum\limits_{\Box} \Big[ \frac{1}{N} Re Tr \Pi_\Box - 1 \Big] \, ,
\end{equation}
where the sum extends over all possible plaquettes $\Box$ and $\Pi_\Box$ 
is the oriented product of link variables around $\Box$; 
for the $U(1)$ gauge theory, the factor $2N/g^2$ is replaced by $1/g^2$.

\begin{figure}
\centering
\includegraphics[width=0.8\textwidth]{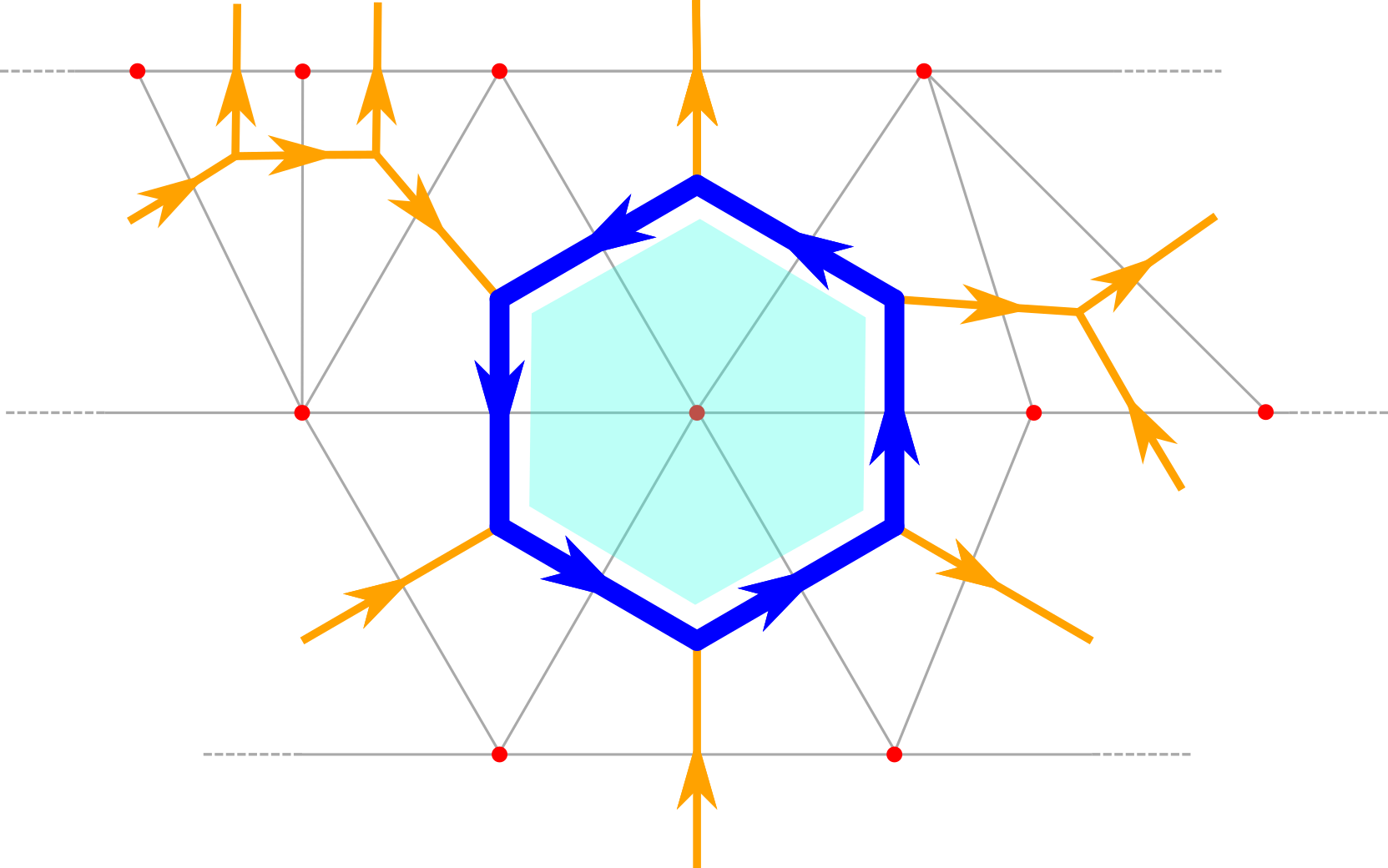}
\caption{Sketch of a plaquette on a section of a typical 2D CDT triangulation with
    vertices (red dots) causally organized in ``foliae'' with Euclidean time 
    increasing vertically.
For simplicity, we have considered a plaquette around a vertex with zero 
curvature, i.e., $n_b = 6$, but, in principle, any number of links from $4$ to $\infty$ is allowed. 
Even if drawn differently, due to the representation on a plane, 
all triangles in this and the following figures 
should be considered equilateral. Taken from~\cite{CDTYM}.}\label{fig:plaquette_draw}
\end{figure}

In order to minimally couple the Yang-Mills theory to gravity, 
it is important to notice that the space of possible configurations for the composite
system is not a simple product space, because the space of gauge field configurations 
depends on the underlying manifold on which it is defined. 
For a fixed manifold $\mcM$, we abstractly denote by $\YMconfsM$ the space of 
all possible gauge field configurations on $\mcM$. 
In terms of (causal) simplicial manifolds $\mcTau$ and gauge configurations $\Phi \in \YMconfs$, 
we can then formally express the Euclidean action for the composite gravity-gauge system as 
$S[\mcTau,\Phi] = S_{CDT}[\mcTau] + S_{\text{YM}}[\mcTau;\Phi]$,
where $S_{CDT}$ is the pure-gravitational part of the action, while 
$S_{\text{YM}}$ represents the minimally coupled action of the field $\Phi$ 
in the backgroud $\mcTau$. 
For the gravitational part, the discretization adopted in CDT is usually the Wick-rotated 
Regge action derived from the continuum Einstein-Hilbert action 
$S_{EH} = \frac{1}{16\pi G}\int \! d^d x \sqrt{-\det(g_{\mu\nu})}(R - 2 \Lambda)$, 
where $\Lambda$ and $G$ are the cosmological and Newton constants respectively and   
$R$ is the scalar curvature. 
Regarding the gauge fields, as in the flat case, 
different discretization choices are possible; we opted for an analogue of 
the plaquette action~\eqref{eq:un-YMact} with link variables associated to the edges of
the graphs \emph{dual} to triangulations (an example in the 2D case is shown in Figure~\ref{fig:plaquette_draw}).
For any dimension $d$, the form of the plaquette action over 
a triangulation $\mcTau$ can be written 
as follows
\begin{equation}\label{eq:un-YMact_cdt}
    S_{\text{YM}}[\mcTau;\Phi] \equiv - \beta\!\!\!\! \sum\limits_{b \in \mcTau^{(d-2)}}\!\! \frac{1}{n_b} \Big\lbrack \frac{1}{N} Re Tr \Pi_b - 1\Big\rbrack,
\end{equation}
where the sum is over all $(d-2)$-simplexes of the triangulation $\mcTau^{(d-2)}$,
$\Pi_b$ is the oriented product of link variables around $b$ (i.e., the plaquette) and
$n_b$ is the plaquette length, which, in turn, is proportional to the area inside the 
plaquette. Notice that $\beta$ is proportional to $1/g^2$ with a non-universal 
factor depending on the discretization and it would in general differ even in the flat case 
for square, hexagonal or kagome lattices.
More details on the choice of the dual lattice instead of the direct one, 
and about the presence of the weight $\frac{1}{n_b}$, 
can be found in the original paper~\cite{CDTYM}. 

\section{Outline of the algorithm in 2D}\label{sec:algo}
While the expression stated above are valid in any dimension, 
here we briefly describe the algorithmic setup by specializing the discussion to the 2D case.

The path-integral over configurations of the composite system can be computed by a Monte Carlo
sampling of the discretized distributions $\exp(-S_{\text{CDT}}^{(\text{2D})} - S_{\text{YM}})$,
where the Yang-Mills action takes the form in eq.~\eqref{eq:un-YMact_cdt},
while the CDT action in two dimensions is simply~\cite{cdt_review12}
\begin{equation}\label{eq:mincoup-CDT2Daction}
    S_{\text{CDT}}^{(\text{2D})}[\mcTau]= \lambda N_2[\mcTau],
\end{equation}
where $N_2[\mcTau]$ is the number of triangles in the triangulation $\mcTau$ 
(i.e., the total volume), 
and $\lambda$ is the only free parameter, which is related to the cosmological constant. 

\begin{figure}
\centering
\includegraphics[width=0.8\textwidth]{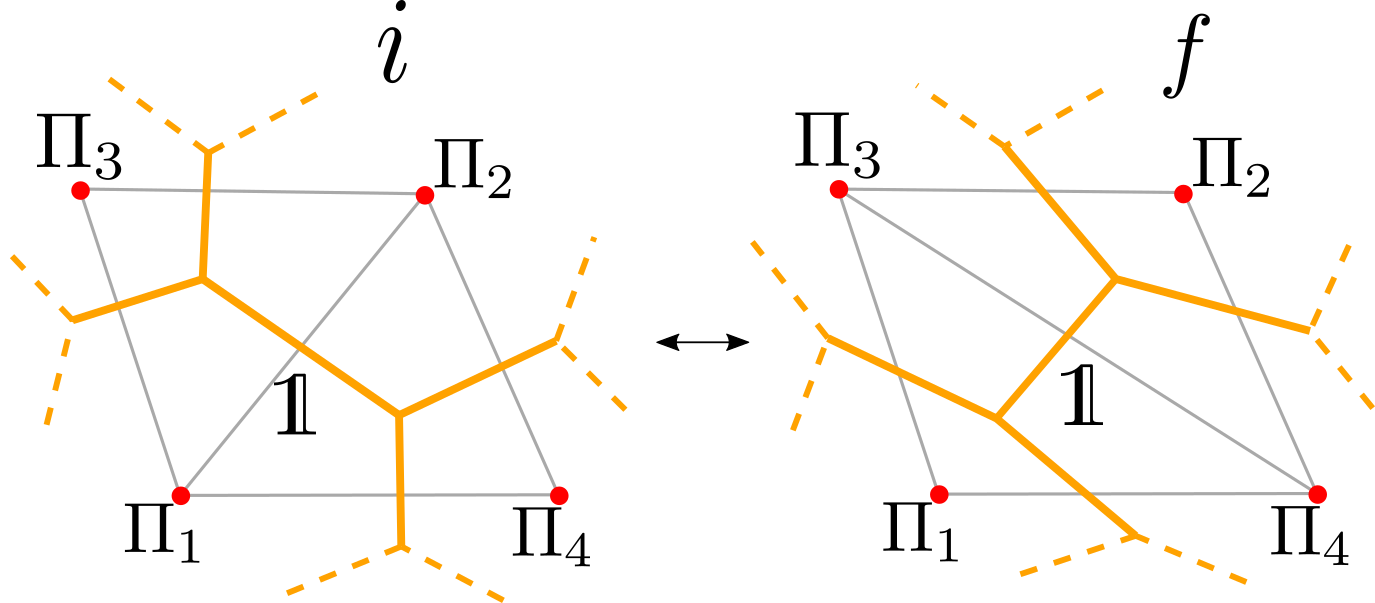}
\caption{Sketch of a $(2,2)$ move. The edge dual to the flipped link is gauge fixed to 
    the identity. Notice that, even if the plaquettes involved $\Pi_{1\leq i \leq 4}$ 
    keep the same value after the move, their contribution to the action does change 
    because the coordination numbers of all 
the four involved vertices change. Taken from~\cite{CDTYM}.}\label{fig:move22_draw}
\end{figure}

\begin{figure}
\centering
\includegraphics[width=0.7\textwidth]{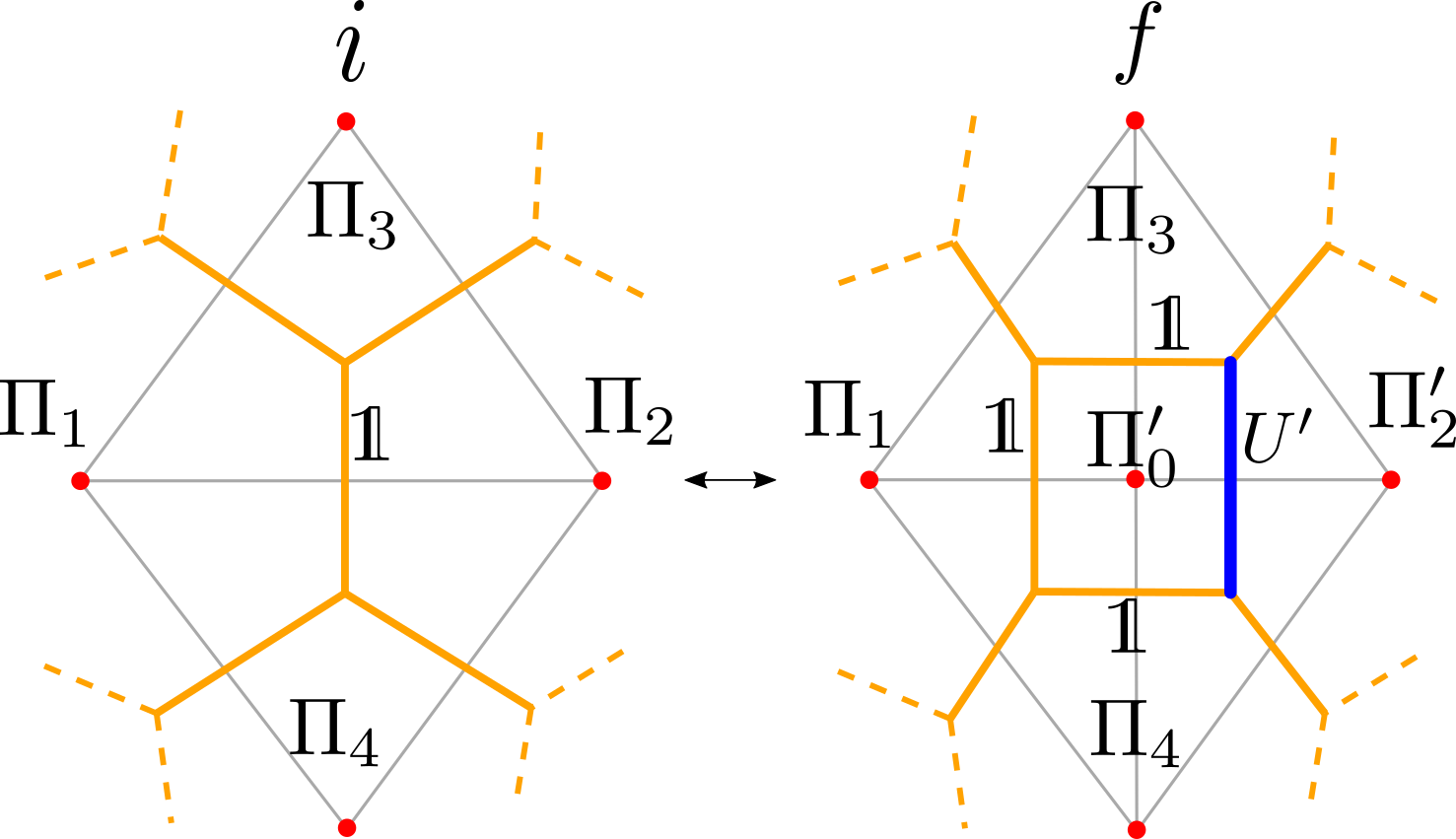}
\caption{Sketch of a $(2,4)$ move and its inverse. The link variable $U^\prime$ (in blue) is the newly extracted one. 
Taken from~\cite{CDTYM}.}\label{fig:move2442_draw}
\end{figure}

In pure gravity CDT, simulations proceed by 
Metropolis{--}Hastings steps~\cite{metro,hastings} which locally change the triangulations
by keeping the causal structure intact: 
the set of causal ergodic moves usually adopted in CDT are called
\emph{Alexander moves}~\cite{alexander,cdt_2dmoves}. 
In two dimensions, these moves are the flip of a time-like link, denoted by $(2,2)$, 
and the creation (or destruction) of a vertex, denoted $(2,4)$ and $(4,2)$, 
where the notation follows the number of triangles involved before and after the move;
these are depicted in figures~\ref{fig:move22_draw} and~\ref{fig:move2442_draw} 
for the underlying triangulation (red dots and grey lines).
In the composite gravity-gauge case, any change in the geometry $\mcTau \to \mcTau^\prime$
would also affect also the space of gauge configurations, 
in such a way that we need to define a map $\YMconfs \to \mathcal{C}_{G}\big(\mcTau^\prime\big)$
between spaces of possibly different numbers of dimensions. 
For example, in the $(2,4)$ move (or its inverse), shown in 
figure~\ref{fig:move2442_draw}, a plaquette is created (or destroyed), 
and this fact reflects non-trivially on the detailed balance condition.
Also the $(2,2)$ move, depicted in figure~\ref{fig:move22_draw}, 
contributes with a non-zero action variation because the $\frac{1}{n_b}$ term in 
equation~\eqref{eq:un-YMact_cdt} changes for all of the four plaquettes involved.

Of course, for moves changing the triangulation, 
the map $\YMconfs \to \mathcal{C}_{G}\big(\mcTau^\prime\big)$ can be chosen arbitrarily, 
but, for performance reasons, it is useful to
keep the acceptance rate high enough by making only minimal local changes to the gauge fields.
Another tool, that proves to be quite useful in the detailed balance analysis, is the freedom
to gauge transform configurations within the same gauge orbit: 
this way, it is possible to \emph{gauge fix} some link variables to the group identity,
greatly simplifying the expressions involved, 
as shown in figures~\ref{fig:move22_draw} and~\ref{fig:move2442_draw}.

\begin{figure}
\centering
\includegraphics[width=0.9\textwidth]{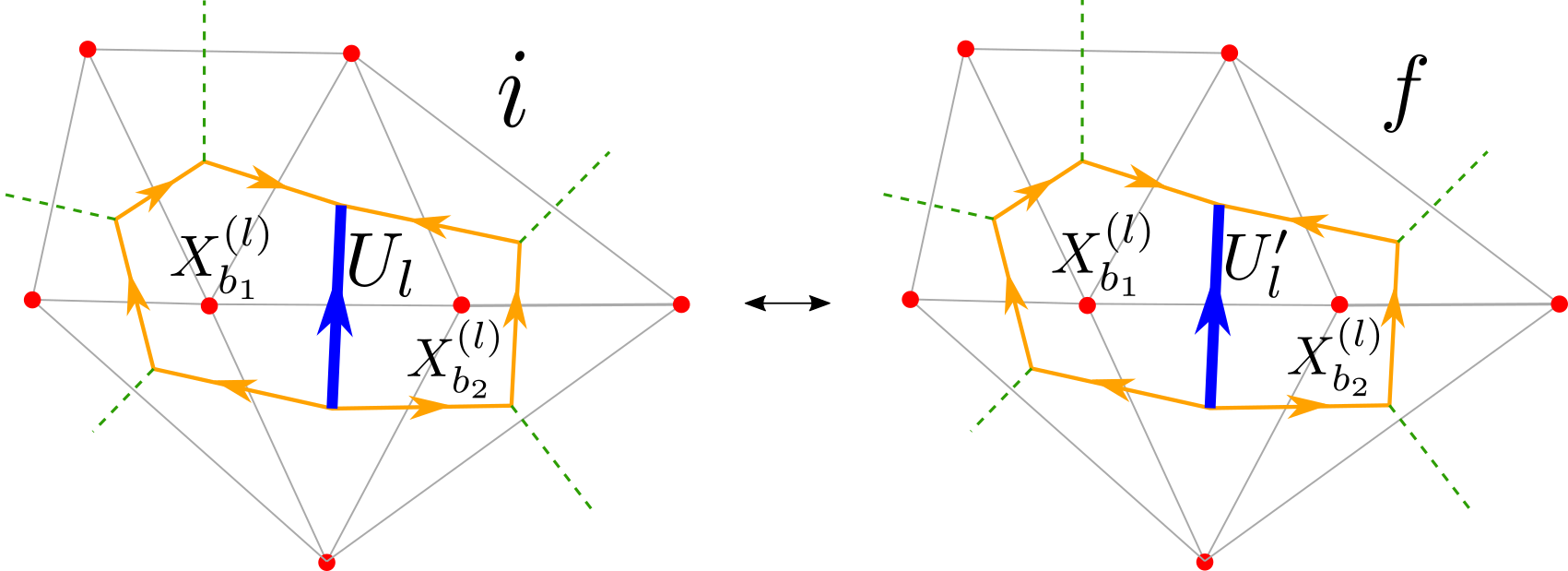}
\caption{Sketch of a typical gauge move. The initial and final state 
    change only in the value of the link variable associated to the edge $l$ (in blue)
    of the graph dual to the triangulation. 
    The staples $X^{(l)}_{b_1}$ and $X^{(l)}_{b_2}$ are built as product of link variables
    oriented as indicated by the arrows (the orange paths). Taken from~\cite{CDTYM}.}\label{fig:movegauge_draw}
\end{figure}

Finally, one has also to guarantee ergodicity within the space 
of gauge field configurations $\YMconfs$ on a fixed triangulation $\mcTau$.
This can be done in analogy to standard Yang{--}Mills simulations,
where one changes just the value of a link variable usually via heat-bath sampling as
in figure~\ref{fig:movegauge_draw},
recalling that staples can have an arbitrary length in this case.

More details on the detailed balance analysis and the sampling strategy 
can be found in the appendix of~\cite{CDTYM}.


\section{Numerical Results}\label{sec:numres}

In this Section we report some of the numerical results obtained in~\cite{CDTYM} 
for $U(1)$ and $SU(2)$ gauge theories minimally coupled to CDT in 2 dimensions.

Before showing the results, we briefly mention the parameters involved in the discretization 
and the numerical setup: 
on the gravity side the bare coupling $\lambda$ acts as a cosmological parameter 
in front of the total volume, since the path-integral is weighted by $\exp(- \lambda V)$; 
in this respect, $\lambda$ can be interpreted as a chemical potential coupled to
the total volume $V$: lowering its value results in an increase of the average volume 
$\langle V \rangle$, up to a certain critical value $\lambda_c$ below which 
the average volume is diverging. For two dimensional CDT, 
this treshold has been computed 
exactly by analytical means and has the value $\lambda_c^{(0)}=\log(2)\simeq 0.693147$
(see Refs.~\cite{two_dim_scaling,two_dim_scaling2,cdt_review12}).
In general, close to the critical point, we expect not only the average volume, but also 
other observables to present a critical behavior which can be fitted to 
a power law scaling of the form:
\begin{equation}\label{eq:un-scalingfunc}
    f(\lambda) = \frac{A}{{(\lambda - \lambda_c)}^\alpha},
\end{equation}
where $\alpha$ is the critical index associated to the observable.
Another parameter is the number of time slices $N_t$, 
which is fixed and constant throughout the simulation 
(since it is kept unchanged by Alexander moves);
we fixed it in such a way that one could neglect finite size effects, 
which could be assessed by measuring the two point correlations functions in the time direction
for either gravity or gauge related observables (these will be introduced in the subsections).
Furthermore, for simplicity, we chose the topology of triangulations to be periodic 
in both the time and space directions\footnote{Notice that Alexander moves also do not change 
the topology, which is then constant during the entire simulation.}.
On the gauge field side, the only parameter is inverse gauge coupling $\beta \propto 1/g^2$, 
defined in eq.~\eqref{eq:un-YMact_cdt}, so that for the composite gravity-gauge theory 
one expects in general some dependence on $\beta$ of the critical value 
$\lambda_c=\lambda_c(\beta)$, below which the average volume diverges.

\subsection{Gravity related observables}

To characterize the gravity part of the system, we have considered two observables:
the total volume of the triangulation $V \equiv N_2$ (i.e., the total number of triangles)
and the volume profiles, i.e., the number of spatial links (spatial edges of the triangles) 
$V(t)\equiv N_{1s}(t)$ in each slice, 
where $t$ labels the slice time\footnote{Actually, the total volume can be computed
from the volume profile due to simplicial constraints in 2D, 
so it doesn't carry new information with respect to the second one.}. 
From the volume profiles one can derive its associated two-point correlation function, 
defined as
\begin{equation}\label{eq:un-corr2prof}
	C_{\text{Vprof.}}(\Delta t) \equiv \frac{\langle V(t) V(t+\Delta t)\rangle - \langle V(t)\rangle \langle V(t + \Delta t) \rangle}{\langle V(t)^2\rangle - \langle V(t)\rangle^2},
\end{equation}
which, at least for $\Delta t \ll N_t$, is expected to decay exponentially as 
$\exp{\big(-\frac{\Delta t}{\xi_{\text{Vprof}}}\big)}$, where 
$\xi_{\text{Vprof}}$ denotes the correlation length of volume profiles.

For $\beta=0$, the Yang-Mills part of the action does not contribute at all, so the
results for gravitational observables are expected to follow exactly the same behavior 
as in the case without gauge fields, $\lambda_c(0)=\log(2)$.
For $\beta>0$, there is some effect of gauge fields on the gravity side; 
for example, a fit of the correlation length of volume profiles $\xi_{\text{Vprof}}$ 
to a shifted power-law behavior (eq.~\eqref{eq:un-scalingfunc})
from simulations with gauge group $U(1)$ at $\beta = 1$
gives the estimate
$\lambda_c^{(G)} =  0.6064(2), \ \nu = 0.53(3)$ 
($\chi^2/{\rm dof} = 8.6/13$). 
\begin{figure}
\centering
\includegraphics[width=0.8\textwidth]{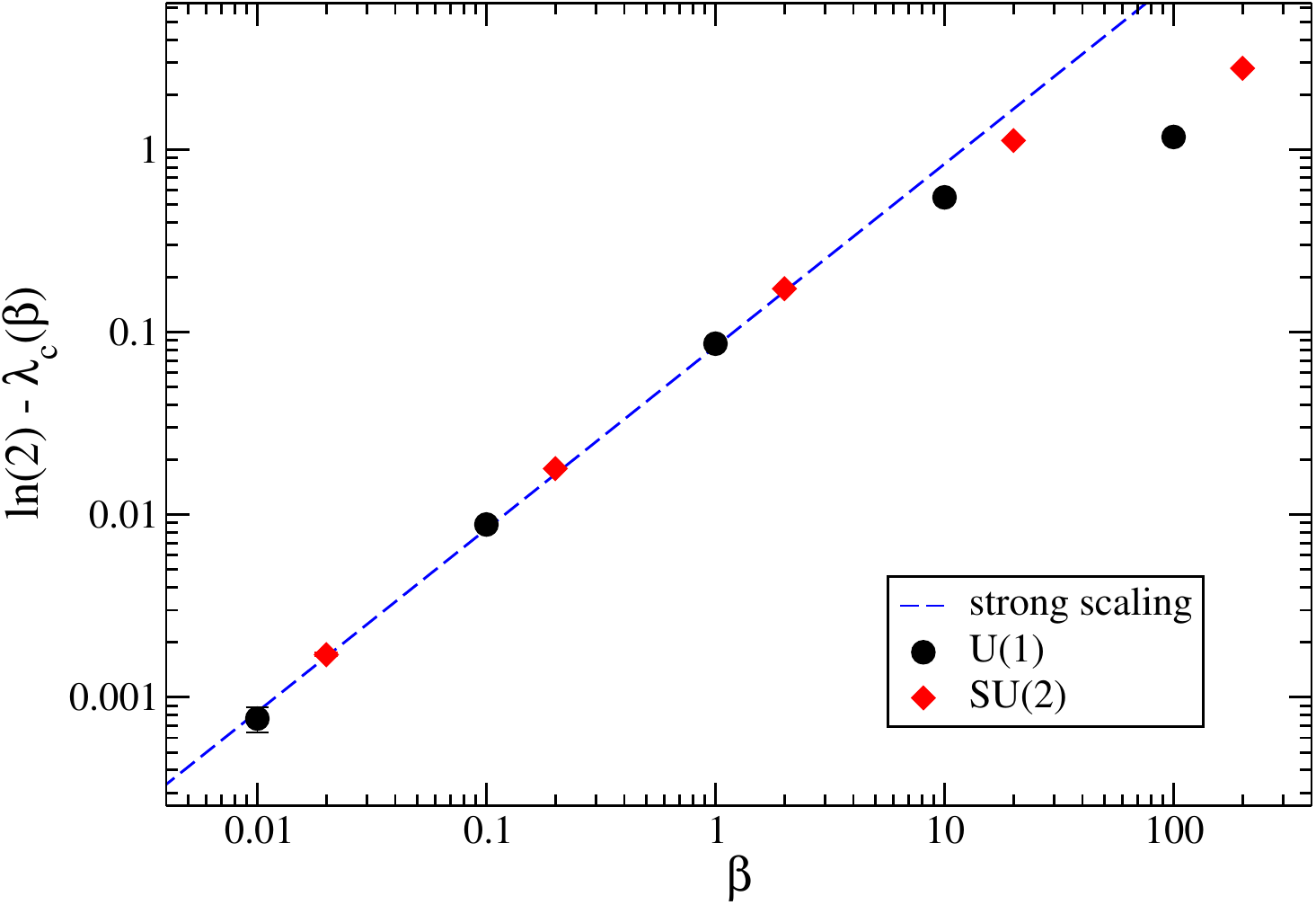}
\caption{Critical points for the $U(1)$ and $SU(2)$ gauge theories minimally coupled to 
    two-dimensional CDT as a function of $\beta$. Taken from~\cite{CDTYM}.
}\label{fig:pseudocritical}
\end{figure}
Notice that the critical value is shifted a slightly below 
the value of the pure gravity theory 
$\lambda_c(0)=\log 2$: 
this is observed for a wide range of $\beta$ considered, 
as shown in figure~\ref{fig:pseudocritical},
where we report the quantity $\log 2 - \lambda_c^{(G)}(\beta)$ 
for the two gauge groups $G=U(1)$ and $G=SU(2)$.
Furthermore, for $\beta \lesssim 1$, the data follows quite well the scaling
$\lambda_c^{(G)}(\beta) =  \log(2) - \beta/12$,
which can be expected from a strong coupling argument (see~\cite{CDTYM}).

\begin{figure}
\centering
\includegraphics[width=0.9\textwidth]{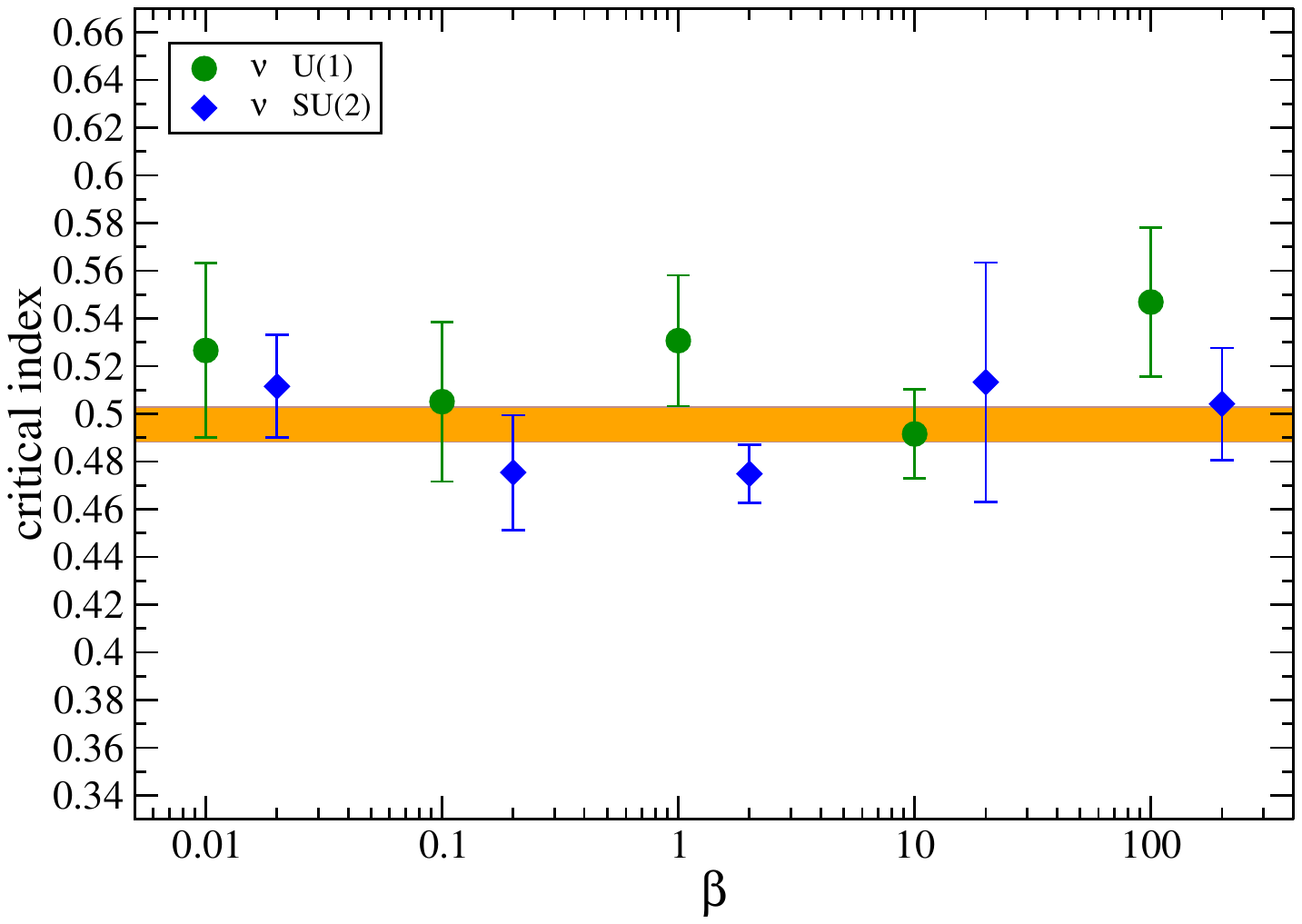}
\caption{Critical index $\nu$ for the correlation length of volume profiles
$\xi_{\text{Vprof.}}$ (see eq.~\eqref{eq:un-scalingfunc}), plotted  
as a function of the bare gauge coupling
$\beta$. 
The horizontal bands represent the results
of a fit to constant values discussed in the text. 
Taken from~\cite{CDTYM}.}\label{fig:fit_critidx}
\end{figure}
Regarding the critical index $\nu$ (associated to $\xi_{\text{Vprof}}$),
figure~\ref{fig:fit_critidx} shows its value for both groups considered 
and for a wide range of the gauge parameter $\beta$:
it is interesting that $\nu$ appears to be independent 
of $\beta$ in the whole range, suggesting that the only effects of the 
gauge fields on gravity observables consist in a shift of 
$\lambda_c$, which, being not observable, can be interpreted just as 
a reparametrization of the bare cosmological parameter.
On the basis of the apparent stability observed,
we have performed a fit of all determinations to a constant
value, obtaining 
$\nu = 0.496(7)$ ($\chi^2 / \textrm{dof} = 9.1/9$),
which is compatible with the value expected in the pure gravity case ($\nu = \frac{1}{2}$)
(see~\cite{two_dim_scaling,two_dim_scaling2,cdt_review12,Zohren:2006tg}).

\subsection{Gauge observables}\label{subsec:gauge_obs}

\begin{figure}
\centering
\includegraphics[width=0.9\textwidth]{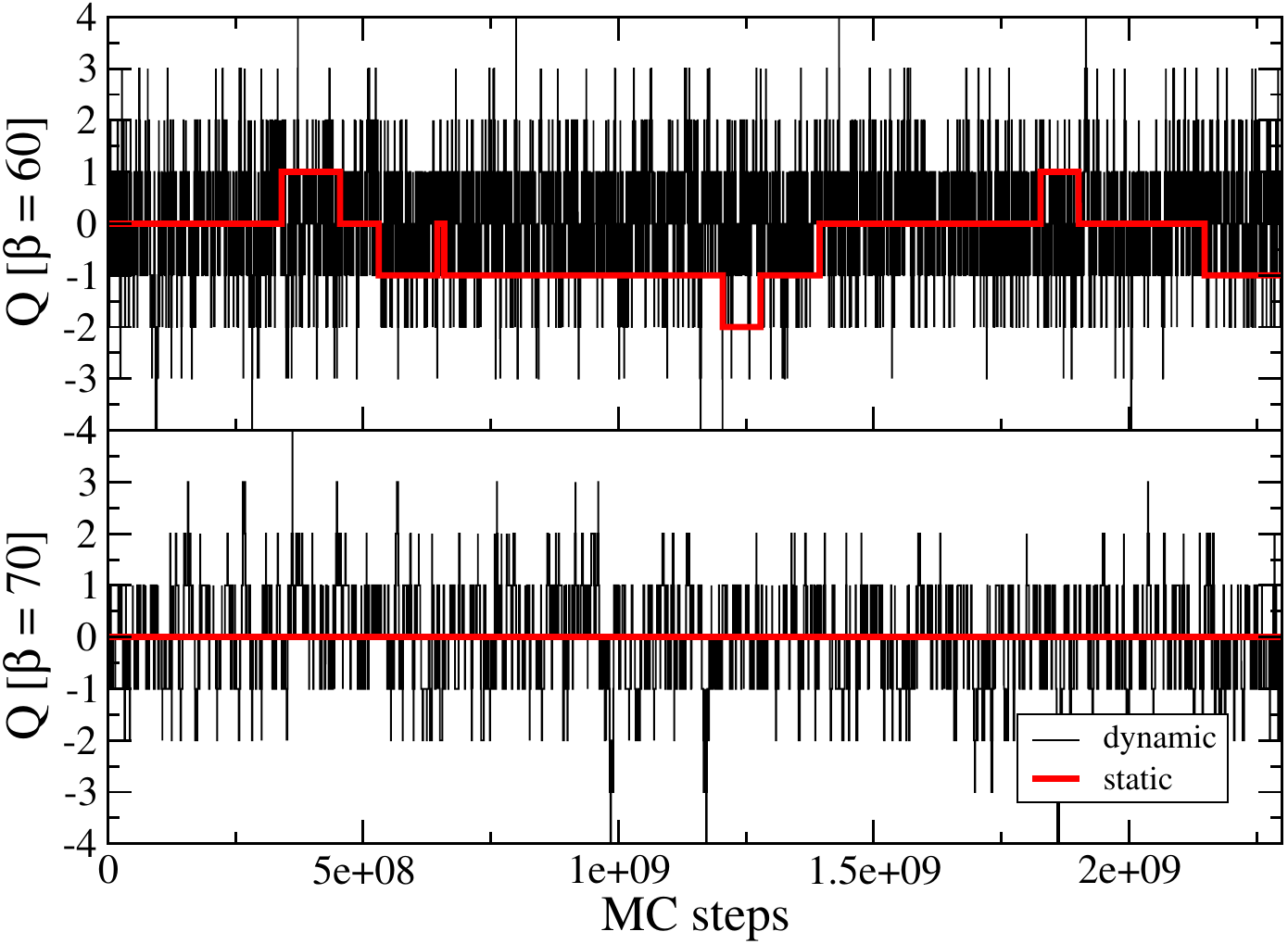}
\caption{Comparison of topological charge histories at $\beta=60$ and $70$
    for the static (flat) and dynamic simulations at average volume $N_2=800$ 
and with $N_t = 20$ slices. Taken from~\cite{CDTYM}.}\label{fig:Qhistories_compare}
\end{figure}
An interesting observable for the gauge part of geometries with toroidal topology
is the topological charge, which acts as a winding number for the gauge field and
is integer-valued only for the $U(1)$ gauge group.
On a 2D lattice, this can be discretized as
\begin{equation}\label{eq:un-topcharge_U1}
	Q \equiv \frac{1}{2 \pi}\sum_{b\in \mcTau^{(0)}} \arg\lbrack Tr (\Pi_b)\rbrack,
\end{equation}
where $\Pi_b$ is the plaquette around the vertex $b$, 
and the argument function $\arg z$ has image in the interval $(-\pi,\pi\rbrack$.
From the distribution of the topological charge, one can derive its cumulants,
which are related to the expansion of the free energy density $f$ of the gauge 
theory in powers of the topological parameter $\theta$. 
In particular the topological susceptibility,
$\chi \equiv \langle Q^2 \rangle /V$, fixes the leading order 
quadratic dependence, $f(\theta) = \chi \theta^2 /2 + O(\theta^4)$. 

A typical behavior of the topological charge for two values of $\beta$ 
($60$ and $70$) is shown in 
Figure~\ref{fig:Qhistories_compare}, where we also compare it with an
analogous simulation on an static flat hexagonal\footnote{We choose the hexagonal lattice
because it maps exactly to the graph dual to triangulated torus without twists.} 
lattice with the same total volume, 
where the infamous \emph{topological freezing}~\cite{Alles:1996vn,top_freeze,Luscher:2011kk,Bonati:2017woi} at large $\beta$s is quite evident.

\begin{figure}
\centering
\includegraphics[width=0.9\textwidth]{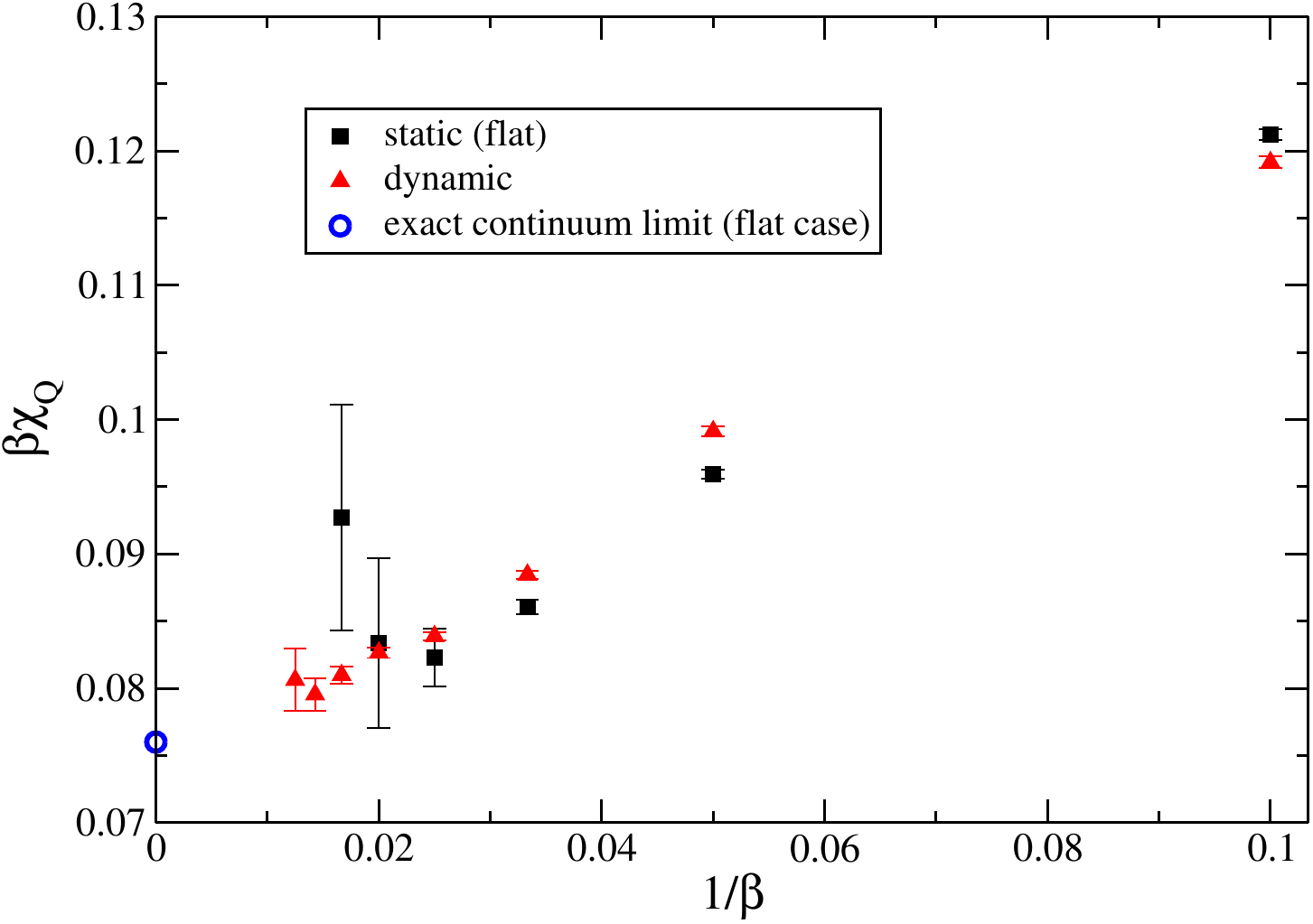}
\caption{Comparison in topological susceptibilities for the static (flat) and dynamic 
simulations at average volume $N_2=800$ and $N_t=20$ slices. Taken from~\cite{CDTYM}.}
\label{fig:topsusc_stat_dyn_20x20}
\end{figure}
This mitigation to the topological freezing in the curved dynamical case,
allows us to obtain an accurate estimation of the susceptibility even at values of $\beta$ 
difficult to reach in the static flat case, as shown in 
figure~\ref{fig:topsusc_stat_dyn_20x20}; here, one can observe a slight difference between
the case with and without gravity, suggesting the presence of some physical effect of gravity
on the gauge fields. 
Furthermore, it is also possible to extrapolate (fitting at $\beta > 30$)
the value of the susceptibility at $\beta\to\infty$, 
which turns out to be $\beta \chi_Q = 0.0758(14)$ 
with a reduced chi-squared $\chi^2/\textrm{dof} = 1.6/3$: 
this appears to be in agreement with the analytical 
prediction for the flat continuum theory~\cite{bonati_flatsuscu1}, which, taking into
account the additional factor 6 in our definition of $\beta$, is $\beta \chi_Q = 3 / (4 \pi^2)$.
However, the limit $\beta\to\infty$ is nai\"ve and not sufficient
in the investigation of the continuum limit of the composite theory, 
which depends also on the $\lambda$ parameter.
As briefly mentioned in the conclusions, 
for a consistent continuum limit analysis, one should introduce another independent scale
besides the correlation length of volume profiles.
To this extent, we also considered
the connected two point correlation functions of
torelonic profiles, defined as:
\begin{equation}\label{eq:un-corr2torel}
	C_{\text{tor}}(\Delta t) \equiv
\frac{\langle Tr \Pi(t) Tr \Pi^\dagger(t+\Delta t) \rangle - 
|\langle Tr \Pi(t) \rangle|^2}{\langle |Tr \Pi(t)|^2\rangle - 
|\langle Tr \Pi(t) \rangle|^2}\,,
\end{equation}
from which one can extract as before the corresponding correlation length $\xi_{\text{tor.}}$.

\section{Discussion and Conclusions}\label{sec:conclusions}

We discussed the numerical implementation of compact gauge fields minimally
coupled to CDT as link variables attached to edges of the graphs dual to triangulations.
In our numerical analysis, we first studied gravity related observables,
namely the total volume and the volume profiles;
while the critical value of the bare cosmological coupling $\lambda$
is $\lambda_c = \log(2)$ in the pure gravity case, the value of $\lambda_c(\beta)$ 
found for the system coupled to gauge fields with a finite bare gauge coupling $\beta$ 
seems to be shifted below $\log(2)$, and, for small $\beta$, 
it follows quite well a strong coupling estimate.
In particular, the critical index $\nu$ associated
with the correlation length of volume profiles
is found to be independent of $\beta$ within errors, and 
a fit to a constant function gives the estimate
$\nu = 0.496(7)$,
i.e.~$\nu$ is compatible with the
analytical prediction obtained in the pure gravity theory. 
These observations suggest that, besides an unobservable shift in its bare parameter, 
gravity is not influenced by gauge fields; this could signal 
that gauge fields may be actually integrated away in this case
as it happens also in other two dimensional contexts (see, e.g., the discussion 
in Refs.~\cite{cdtgauge_anal,cdtgauge_noncompact,Cao:2013na,bonati_flatsuscu1}).

On the other hand, the gauge fields dynamics seems to be affected by the coupling with 
gravity in a less trivial way. In the case of the $U(1)$ gauge group in 2D, we
can study the dependence of the free energy on the topological parameter $\theta$ 
via measurements of the integer valued topological charge. 
For generic gauge groups, an interesting quantity to measure is the torelon correlator 
and its associated correlation length in slice time, which is physically independent
from the correlation length of volume profiles and could therefore be used as
a distinct length scale in the analysis of the continuum limit.

Indeed, as future developments to this exploratory study, 
we plan to investigate, in the region of the $\lambda-\beta$ plane close to the critical line, 
the so-called curves of constant physics, along which the ratio between 
the two types of correlation lengths is fixed to a value 
that, in principle, could identify different physical theories in the continuum limit.
Moreover, in the future we also plan to consider 
higher dimensions and coupling fermionic variables, 
both of which pose some algorithmic challenges.

\acknowledgments{}
We thank Claudio Bonati, Paolo Rossi and Francesco Sanfilippo 
for useful discussions. 
Numerical simulations have been performed on the MARCONI machine at CINECA, 
based on the agreement between INFN and CINECA 
(under projects INF19\_npqcd and INF20\_npqcd), 
and at the IT Center of the Pisa University.

\end{document}